\documentclass[twocolumn,showpacs,preprintnumbers,amsmath,amssymb]{revtex4}
\usepackage{graphicx}
\usepackage{dcolumn}
\usepackage{bm}
\usepackage{bbm}
\usepackage{graphicx,epsf,epsfig}
\usepackage{amsmath}
\usepackage{amssymb}

\usepackage{tabls,multirow}

\def\Tr{\mbox{Tr}\,}

\def\hbar{\hspace{0pt}\raisebox{1pt}{$-$} \hspace{-7pt} h}

\newcommand{\be}{\begin{equation}}
\newcommand{\ee}{\end{equation}}
\newcommand{\bd}{\begin{displaymath}}
\newcommand{\ed}{\end{displaymath}}
\newcommand{\bea}{\begin{eqnarray}}
\newcommand{\eea}{\end{eqnarray}}
\newcommand{\nn}{\nonumber}

\def\so10{$SO(10)$}

\begin{document}
\title{Third Family Corrections to Tri-bimaximal Lepton Mixing and a New Sum Rule}
\date{\today}
\author{Stefan Antusch} \email{antusch@mppmu.mpg.de}
\affiliation{Max-Planck-Institut f\"ur Physik (Werner-Heisenberg-Institut)
F\"ohringer Ring 6, D-80805 M\"unchen, Germany}
\author{Stephen F. King}\email{sfk@hep.phys.soton.ac.uk}
\affiliation{School of Physics and Astronomy, University of Southampton,
SO16 1BJ Southampton, United Kingdom}
\author{Michal Malinsk\'{y}}\email{malinsky@phys.soton.ac.uk}
\affiliation{School of Physics and Astronomy, University of Southampton,
SO16 1BJ Southampton, United Kingdom}
\begin{abstract}
We investigate the theoretical stability of the
predictions of tri-bimaximal neutrino mixing with respect to
third family wave-function corrections.
Such third family wave-function corrections
can arise from either the canonical normalisation of the
kinetic terms or renormalisation group running effects. 
At leading order both sorts of corrections
can be subsumed into a single universal parameter.
For hierarchical neutrinos, 
this leads to a new testable lepton mixing sum rule $s = r \cos \delta +
\frac{2}{3} a$ (where $s,r,a$ describe 
the deviations of solar, reactor and atmospheric mixing angles
from their tri-bimaximal values, and $\delta$ is the observable Dirac
CP phase) which is stable under all leading order third family 
wave-function corrections, as well as Cabibbo-like charged lepton 
mixing effects. 
\end{abstract}
\maketitle
\section{Introduction}
Since the discovery of neutrino masses and large lepton mixing angles,
the flavour problem of Standard Model (SM) has received much
attention. As the precision of the neutrino data has improved, it has
become apparent that lepton mixing is consistent with the so called
Tri-bimaximal (TB) mixing pattern \cite{HPS},
\be\label{UTB}
U_{\mathrm{TB}}=
\begin{pmatrix}
\sqrt{\tfrac{2}{3}} & \tfrac{1}{\sqrt{3}} & 0 \\
- \tfrac{1}{\sqrt{6}}  & \tfrac{1}{\sqrt{3}} &  \tfrac{1}{\sqrt{2}} \\
 \tfrac{1}{\sqrt{6}} & - \tfrac{1}{\sqrt{3}} &  \tfrac{1}{\sqrt{2}}
\end{pmatrix}.\,P_{\mathrm{M}}
\ee
where
$P_{\mathrm{M}}$
is the so far experimentally undetermined diagonal phase matrix encoding the two
observable Majorana phase differences. 
Many models attempt
to reproduce this as a theoretical prediction
\cite{sumrule,Frampton:2004ud,Altarelli:2006kg,Ma:2007wu,deMedeirosVarzielas:2005ax,Harrison:2003aw,Chan:2007ng,Antusch:2007jd}.
Since the forthcoming neutrino experiments will be sensitive to
small deviations from TB mixing, it is important to quantify
the ``theoretical'' uncertainty inherent in such TB mixing
predictions.

In many classes of models 
TB mixing arises purely from the neutrino sector \cite{King:2006hn},
subject to deviations due to charged lepton
sector corrections \cite{sumrule,Frampton:2004ud}.
If these charged lepton corrections are ``Cabibbo-like'' in nature 
(i.e.\ dominated by a 1-2 mixing), 
it leads to a predictive sum rule \cite{sumrule} 
which may be expressed in terms
of the parametrisation in \cite{King:2007pr} as $s=r\cos
\delta$, where $s$ and $r$ describe 
the deviations of solar and reactor mixing angles
from their tri-bimaximal values, and $\delta$ is the observable Dirac
CP phase in the standard  
parameterisation \cite{Yao:2006px}. 

Another source of theoretical uncertainty in TB mixing schemes is the
renormalisation group (RG) running \cite{RGE} of the relevant
quantities from the high energy (usually the unification scale
$M_{\mathrm{G}}$), where the theory is defined, to the
electroweak scale $M_{Z}$ appropriate for experimental measurements.
The dominant source of RG corrections to lepton mixing arises typically from 
the large tau lepton and third family neutrino Yukawa couplings,
leading to relatively large wave-function corrections in the
framework of supersymmetric models. Such RG corrections
can be readily estimated analytically \cite{RGEanalytical,Dighe:2006sr}
for the TB mixing case with hierarchical light neutrinos 
considered here. Diagrammatically, such RG
corrections correspond to loop diagrams involving third family
matter and Higgs fields and their superpartners. Although suppressed
by the loop factor of ${1}/{16\pi^2}$, they can be relevant since the
loop factor is multiplied by a large logarithm of the ratio of energy
scales.

Apart from RG effects there is another type of third family
wave-function correction which emerges at tree-level in certain
classes of models, and thus can potentially be rather large.  These
corrections modify the kinetic terms in the Lagrangian causing them to
deviate from the standard (or canonical) form. Before the theory can
be reliably interpreted, field transformations must be performed in
order to return the kinetic terms back to canonical form which,
however, leads to appropriate modifications of the Yukawa
couplings. It is interesting that these effects are largest in many of
the theories that predict TB mixing, especially those based on 
non-Abelian family symmetries spanning all three families of SM
matter (see e.g.\ \cite{King:2003xq,King:2004tx,Antusch:2007re}). In
such models the canonical normalisation (CN) corrections can in
certain cases even exceed the effects due to RG running.

In this Letter we shall provide a unified treatment of all the above
sources of theoretical corrections to the TB mixing mixing, namely due
to: {\it i}\,) RG corrections, {\it ii}\,) CN
corrections and {\it iii}\,) charged lepton corrections.  We will
present a novel testable neutrino mixing sum rule which, at leading
order, is stable under all these effects.

\section{General formalism}
Suppose the original (before the RG and CN corrections
are accounted for) charged lepton ($l$) and Majorana neutrino mass
matrices $\hat M_{l}$ and $\hat m_{\nu}$ are diagonalised by means of
unitary transformations $\hat V_{L}^{l}\hat M_{l}\hat
V_{R}^{l\dagger}=\hat M_{l}^{D}$ (we shall work in LR chirality basis)
and $ \hat V_{L}^{\nu}\hat m_{\nu}\hat V_{L}^{\nu T}=\hat m_{\nu}^{D}
$ so that the uncorrected lepton mixing matrix 
obeys $\hat U_{\mathrm{PMNS}}=\hat V_{L}^{l}\hat
V_{L}^{\nu\dagger}$.

The effect of both CN  and leading logarithmic RG corrections 
on the $\hat M_{l}$ and $\hat m_{\nu}$ matrices
can be described by
a pair of transformation matrices $P_{L,R}$ 
\be\label{cannormmatrices} \hat M_{l}\to P_L^T\hat
M_{l}P_R\equiv M_{l}, \quad \hat
m_{\nu}\to P_{L}^{T}\hat m_{\nu}P_{L}\equiv m_{\nu} \ee 
which induce
a relevant change on $\hat V_{L,R}^{l}\to V_{L,R}^{l}$ and $\hat
V_{L}^{\nu}\to V_{L}^{\nu}$ so that $
V_{L}^{l}M_{l}V_{R}^{l\dagger}=M_{l}^{D}$ and $
V_{L}^{\nu}m_{\nu}V_{L}^{\nu T}=m_{\nu}^{D} $ 
 and thus
 $U_{\mathrm{PMNS}}=V_{L}^{l}V_{L}^{\nu\dagger}$
is the {\it physical} lepton mixing matrix (after global
rephasing). 
One can always write 
$P_{L,R}=p_{L,R}(\mathbbm{1}+\Delta P_{L,R})$ where, as we shall see, the constants $p_{L,R}$ have no effect on the mixing angles and
$\Delta P_{L,R}$
denote the corrections from the flavour non-universal part of the RG and CN effects to be identified later.
Equation (\ref{cannormmatrices}) then implies
\bea\label{basicrelations2} p_{L}p_{R}V_{L}^{l}(\mathbbm{1}+\Delta
P_{L}^{T})\hat M_{l}(\mathbbm{1}+\Delta P_{R})V_{R}^{l\dagger} & =
& M_{l}^{D}\nn \;, \\ 
p_{L}^{2}V_{L}^{\nu}(\mathbbm{1}+\Delta P_{L}^{T})\hat
m_{\nu}(\mathbbm{1}+\Delta P_{L})V_{L}^{\nu T}&=& m_{\nu}^{D}\;. \eea 
If all the physical spectra are sufficiently
hierarchical, the smallness of $\Delta P_{L,R}$
factors ensures only small differences between $\hat V_{L}^{f}$ and $V_{L}^{f}$ (for $f=l,\nu$), in particular  
\be\label{mixingchange}
V_{L}^{f}=W_{L}^{f} \hat V_{L}^{f}=e^{i\Delta W_{L}^f} \hat V_{L}^{f}
\ee 
where $W_{L}^{f}$ are
small unitary rotations in the unity neighborhood with $\Delta
W_{L}^{f}$ denoting their Hermitean generators. 
One can disentangle the
left-handed and right-handed rotations in the charged lepton formula
in (\ref{basicrelations2}) 
by considering $M_{l}M_{l}^{\dagger}$ with the result 
\footnote{Notice that these formulae provide only the off-diagonal entries of
$\Delta W^{f}_{L}$'s. Indeed, formulae (\ref{basicrelations2}) and (\ref{mixingchange}) leave three
phases of the diagonalization matrices $W_{L}^{l,\nu}$ unconstrained so we shall conventionally put $(\Delta W_{L}^{l,\nu})_{ii}=0$.}
\bea
(\Delta W_L^{l})_{ij}^{i\neq j}&\approx&\frac{i}{\hat m_{j}^{l2}-\hat
m_{i}^{l2}}\!\! \left[ (\hat m_{i}^{l2}+\hat m_{j}^{l2})\!\left(\hat
V_{L}^{l}\Delta P_{L}^{T}\hat V_{L}^{l\dagger}\right)_{ij}\right.\nn\\
&+ &\left.2\hat
m_{i}^{l}\hat m_{j}^{l}\left(\hat V_{R}^{\nu}\Delta P_{R}\hat
V_{R}^{l\dagger}\right)_{ij} \right],  \label{DeltaWl}
\eea
where the eigenvalues $\hat m_{i}^{l2}$ of
the original $\hat M_{l}$ matrix can, at leading order, be 
identified with the physical charged lepton masses.  Similarly, the
neutrino sector corrections obey (replacing $\hat M_{l}\to \hat
m_{\nu}$, $V_{L}^{l}\to V_{L}^{\nu}$, $V_{R}^{l}\to V_{L}^{\nu*}$ and
$\Delta P_{R}\to \Delta P_{L}$ in formula (\ref{DeltaWl}) above) 
\bea
(\Delta W_L^{\nu})_{ij}^{i\neq j}&\approx&\frac{i}{\hat m_{j}^{\nu2}-\hat
m_{i}^{\nu2}}\!\! \left[ (\hat m_{i}^{\nu2}+\hat m_{j}^{\nu2})\!\left(\hat
V_{L}^{\nu}\Delta P_{L}^{T}\hat V_{L}^{\nu\dagger}\right)_{ij}\right.\nn\\
&+ &\left.2\hat
m_{i}^{\nu}\hat m_{j}^{\nu}\left(\hat V_{L}^{\nu}\Delta P_{L}^{T}\hat
V_{L}^{\nu\dagger}\right)_{ji} \right].  \label{DeltaWnu}
\eea 

From equations (\ref{mixingchange}), (\ref{DeltaWl}) and (\ref{DeltaWnu}) 
one can write the corrected
(i.e.\ {\it physical}\,) lepton
mixing matrix $U_{\mathrm{PMNS}}=V_{L}^{l}V_{L}^{\nu\dagger}$
in terms of the original $\hat U_{\mathrm{PMNS}}$ as 
$U_{\mathrm{PMNS}}=
\hat U_{\mathrm{PMNS}}+\Delta
U_{\mathrm{PMNS}} \nn 
$
where
\be\label{MNSchange} \Delta
U_{\mathrm{PMNS}}\approx i\left(\Delta W_{L}^{l}\hat U_{\mathrm{PMNS}}-\hat U_{\mathrm{PMNS}}\Delta
W_{L}^{\nu\dagger}\right). 
\ee

Due to the assumed hierarchy in the physical spectra, the first
terms in equations (\ref{DeltaWl}) and (\ref{DeltaWnu})
dominate over the second (thus avoiding the ambiguity
in the unknown structure of the right-handed (RH) rotations in the charged
lepton sector) and so we shall neglect the latter and 
focus on the left-handed (LH) sector.

The RG effects in the supersymmetric case yield at leading order \cite{RGE} $P^{RG}_{L}=r_{L}\mathbbm{1}+\Delta P_{L}^{RG}$ where
\be\label{rL}
r_{L}=1-\frac{1}{16\pi^{2}}\left[3\!\left(\Tr Y_{u}^{\dagger}Y_{u}\!-\!\hat g^{2}\right)\ln \frac{M_{\mathrm{G}}}{M_{Z}}\!+\!\Tr Y_{\nu}^{\dagger}Y_{\nu}\ln \frac{M_{\mathrm{G}}}{M_{N}}\right]
\ee
(with $\hat g^{2}\equiv g_{2}^{2}+\tfrac{1}{5}g_{1}^{2}$) accounts for flavour-universal contribution 
while 
\be\label{DeltaRG}
\Delta P_L^{RG} = - \frac{1}{16 \pi^2}\left[Y_{l}^{*}Y_{l}^{T}  \ln \frac{M_{\mathrm{G}}}{M_{Z}}
+ Y_{\nu}^{*} Y_{\nu}^{T} \ln \frac{M_{\mathrm{G}}}{M_{N}}\right],
\ee
denotes the flavour non-trivial piece. In (\ref{rL}) and (\ref{DeltaRG}), $M_{N}$ denotes the mass of the lightest RH neutrino.
Notice that since $r_{L}$ is close to 1 one can write at leading order 
\be
P^{RG}_{L}\approx r_{L}(\mathbbm{1}+\Delta P_{L}^{RG})
\ee 
rendering the $r_{L}$ factor irrelevant for the mixing angles. 


Turning to CN effects, with the non-canonical LH lepton doublet ($L$)
kinetic term written as $iL^{\dagger}/\!\!\!\!DK_LL$ and $K_L=k_{L}(\mathbbm{1}+\Delta K_L)$,
the CN transformation can be written in the form of (\ref{cannormmatrices}) with
(c.f.~\cite{King:2003xq,King:2004tx,Antusch:2007re})
\be\label{KL}
P_L^{CN}=(\mathbbm{1}+\Delta P_L^{CN})/\sqrt{k_{L}}\;\;\mathrm{where}\;\; \Delta P^{CN}_{L}\approx-\tfrac{1}{2}\Delta K_{L}.
\ee
At leading order, $P_{L}^{CN}$ fulfills $(P_L^{CN})^{-1 \dagger} (P_L^{CN})^{-1}=K_L$.

Finally, one can combine both RG and CN effects under a single transformation satisfying at leading order
\be
P_{L}\approx P_{L}^{CN}P_{L}^{RG}\approx \tfrac{r_{L}}{\sqrt{k_{L}}}(\mathbbm{1}+\Delta P_{L}^{CN}+\Delta P^{RG}_{L})
\ee
yielding the assumed form of $P_{L}=p_{L}(\mathbbm{1}+\Delta P_{L})$ with $p_{L}=r_{L}/\sqrt{k_{L}}$ and 
$\Delta P_{L}=\Delta P_{L}^{RG}+\Delta P_{L}^{CN}$.

Using these results,
the leading order 
corrections to lepton mixing from RG and CN effects can be 
calculated.


\section{Third family corrections to tri-bimaximal lepton mixing}
In this section we shall apply the formalism of
section II to TB lepton mixing, where we assume to begin with that 
this mixing originates entirely from the neutrino sector 
and subsequently extend the analysis to include
corrections from charged lepton mixing \cite{King:2006hn}. 
Thus, let us assume first
that the lepton mixing predicted by some underlying theory in the
absence of RG and CN corrections happens to be exactly tri-bimaximal
$\hat U_{\mathrm{PMNS}}=\hat V_{L}^{l}\hat V_{L}^{\nu\dagger}= U_{\mathrm{TB}}$
where $\hat V_{L}^{l}= \mathbbm{1}$ while $\hat V_{L}^{\nu\dagger}=
U_{\mathrm{TB}}$.  Including RG and CN corrections, $\hat V_{L}^{l}$ and $\hat V_{L}^{\nu\dagger}$ then change according to
(\ref{mixingchange}) with the
correction matrices given by equations (\ref{DeltaWl}), (\ref{DeltaWnu}).

Here we shall restrict ourselves to the dominant third family
wavefunction corrections, so that from (\ref{DeltaRG}) one can
write $\Delta P_L^{RG}=-\tfrac{1}{2}\mathrm{diag}(0,0, \eta^{RG})$
with $\eta^{RG}=\left[y_\tau^2\ln
(M_{\mathrm{G}}/M_Z)+y_{\nu_3}^2\ln
(M_{\mathrm{G}}/M_{N})\right]/{8 \pi^2}$.  If the third family
effects dominate also the form of $K_L$ in equation (\ref{KL}),  
 \ $\Delta K_L$ is
a matrix controlled by the 33 entry and $\Delta
P_L^{CN}=-\tfrac{1}{2}\mathrm{diag}(0,0, \eta^{CN})$ with
$\eta^{CN}={(\Delta K_L)_{33}}$. Then at the leading order $\Delta
P_L=-\tfrac{1}{2}\mathrm{diag}(0,0, \eta)$ is governed by a {\it
universal} parameter $\eta=\eta^{RG}+\eta^{CN}$.  
This is the case, for instance, in all
models in which a family symmetry spans all three SM matter families
\cite{King:2003xq,King:2004tx,Antusch:2007re}. In such theories
the third family corrections to the kinetic function $K_{L}$ 
have the same origin as the third family Yukawa couplings,
and $\eta^{CN}$ can be as large as $y_{\tau}^{2}$ (in the $SO(3)$-type of
models) or $y_{\tau}$ (for underlying $SU(3)$ flavour
symmetry). However, it is also possible that $\eta^{CN}\ll y_{\tau}$ in
certain classes of models, and the effect is strongly dependent on the
details of the underlying 
theory \cite{deMedeirosVarzielas:2005ax,inpreparation}.

Given the diagonal form of $\Delta P_{L}$, and our assumption that
$\hat V_{L}^{l}= \mathbbm{1}$, one gets from equation (\ref{DeltaWl})
$\Delta W_{L}^{l}=0$ and thus $V_{L}^{l}=\mathbbm{1}$ so the corrected lepton
mixing matrix is given by just
$V_{L}^{\nu\dagger}=U_{\mathrm{TB}}+\Delta U_{\mathrm{TB}}$ with
$\Delta U_{\mathrm{TB}}=-iU_{\mathrm{TB}}\Delta W_{L}^{\nu\dagger}$
from equation (\ref{MNSchange}).  If the neutrino spectrum is hierarchical, the
leading (first) term in (\ref{DeltaWnu}) yields
\be\label{universalproportionality} \Delta
(U_{\mathrm{TB}})_{ij}^{i<j} \approx -
(U_{\mathrm{TB}})_{i\underline{j}}\left(U_{\mathrm{TB}}^{\dagger}\Delta
P_{L}^{T}U_{\mathrm{TB}}\right)_{\underline{j}\underline{j}}\;\;
\mathrm{(no\;} j\; \mathrm{ sum.).}  \ee We can see that, at the
leading order, the corrections to $(U_{\mathrm{TB}})_{ij}$ for ${i<j}
$ are proportional to $(U_{\mathrm{TB}})_{ij}$ itself, and thus for
example both RG and CN corrections to $\theta_{13}$ are zero due to
$(U_{\mathrm{TB}})_{13}=0$
\footnote{Note also that 
 (\ref{universalproportionality}) implies that 
the Majorana phases have no effect on the corrections to the 
mixings angles.
%
}.  
However the result $\Delta
(U_{\mathrm{TB}})_{13}=0$ arises only at leading order in
small quantities ($\eta$ and $\sqrt{{\Delta m^{2}_{21}}/{\Delta
m^{2}_{31}}} \approx m_2/m_3 \approx 1/5$) and gets lifted at
next-to-leading level. Restoring the second term in equation
(\ref{DeltaWnu}) one obtains:
\be\label{13correction} \theta_{13}\approx
\tfrac{1}{3\sqrt{2}}|\eta|\sqrt{{\Delta m^{2}_{21}}/{\Delta
m^{2}_{31}}}\approx 4\times 10^{-2}|\eta|\;.  \ee
\begin{figure}[t]
\centering
$\ensuremath{\vcenter{\hbox{\includegraphics[width=8cm]{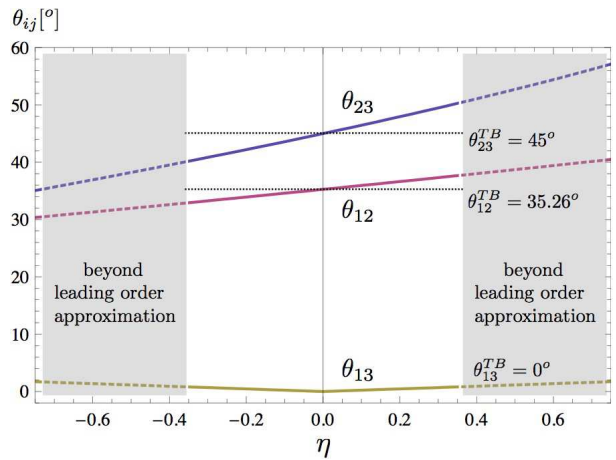}}}}$ 
 \caption{\label{fig:corrections-theory} 
Renormalisation group and canonical normalization corrections to tri-bimaximal neutrino mixing. The shaded regions correspond to $\eta$-values outside the linear approximation.}
\end{figure}
All together, this yields at the leading order
\be
V_L^{\nu\dagger}\approx  \label{correctionTBcase}\begin{pmatrix}
\sqrt{\tfrac{2}{3}}(1-\tfrac{1}{12}\eta) & \tfrac{1}{\sqrt{3}}(1+\tfrac{1}{6}\eta) & 0.04\times |\eta|e^{-i\delta}\\
- \tfrac{1}{\sqrt{6}}(1-\tfrac{1}{12}\eta)  & \tfrac{1}{\sqrt{3}}(1-\tfrac{1}{3}\eta) &  \tfrac{1}{\sqrt{2}}(1+\tfrac{1}{4}\eta) \\
 \tfrac{1}{\sqrt{6}}(1+\tfrac{5}{12}\eta) & - \tfrac{1}{\sqrt{3}}(1+\tfrac{1}{6}\eta) &  \tfrac{1}{\sqrt{2}}(1-\tfrac{1}{4}\eta)
\end{pmatrix}
\ee which is unitary up to ${\cal O}(\eta^{2})$.  In terms of
the deviations from the exact TB mixing parametrized
\cite{King:2007pr} by $\sin \theta_{12}=(1+s)/\sqrt{3}$, $\sin
\theta_{23}=(1+a)/\sqrt{2}$ and $\sin \theta_{13}=r/\sqrt{2}$ one gets
(so far without including charged lepton corrections): \be\label{rsa} r\approx
6\times 10^{-2} |\eta|,\quad s=\tfrac{1}{6}\eta\quad \mathrm{and}\quad
a=\tfrac{1}{4}\eta.  \ee We see that in particular $\theta_{13}$ is
rather stable and the atmospheric $\theta_{23}$ is changing faster
with $\eta$ than the solar $\theta_{12}$,
as shown in FIG. \ref{fig:corrections-theory}.

In any realistic model, the charged lepton mixing corrections entering
$U_{\mathrm{PMNS}}$ must be taken into account. It is well known that
if $\hat V_{L}^{l}$ is Cabibbo-like
with $\theta_{12}^{l}$ being the only non-negligible mixing
angle, then (ignoring RG and CN effects)
this gives rise to a particular pattern of corrections to
$\theta_{13}$ and $\theta_{12}$ that obey (at the high scale) the relation
$s=r\cos\delta$ with, e.g., $r\approx \theta_C/3$ for $\theta_{12}^{l}\approx \theta_C/3$ 
($\theta_{C}$ is the Cabibbo angle) in many unified models \cite{sumrule}.

We can include the
leading order RG and CN corrections to $s=r\cos\delta$ by considering
only the neutrino sector effects (encoded in $\Delta W_{L}^{\nu}$) for
the individual terms. (This follows since
for the Cabibbo-like $\hat V_{L}^{l}$ and 
$\Delta P_{L}$ dominated by the 33 entry, 
it is still the case that $\Delta W_{L}^{l}=0$.)
Therefore the previous example provides a
good estimate of the relevant corrections at leading order in
small quantities (including now also the charged lepton mixing
$\theta_{12}^{l}$). Neglecting the subleading correction in 
(\ref{13correction}), from equation (\ref{rsa}) one obtains 
$s=r\cos\delta+\tfrac{1}{6}\eta$,
which can be rewritten in terms of only measurable
quantities in form of a new sum rule \be\label{sumrule}
s=r\cos\delta+\tfrac{2}{3}a \; .  \ee 

The new sum rule in equation (\ref{sumrule}) 
is stable under the considered
theoretical corrections and additionally involves the 
deviation of atmospheric mixing from maximality \cite{Antusch:2004yx}.
The main sources of remaining uncertainties in
formula (\ref{sumrule}) are the neglected (order $4\% 
|\eta|$)
corrections to $r\cos\delta$ due to the subleading contribution
(\ref{13correction}), the higher order corrections to 
$\Delta P_{L}^{RG}$ and
$\Delta P_{L}^{CN}$ (all suppressed by the relevant Yukawa coupling
ratios) and the higher order $\eta$-effects. A rigorous derivation and
detailed discussion of formula (\ref{sumrule}) will be given in a
forthcoming paper \cite{inpreparation}.


\section{Conclusions}

We have presented a unified formalism for 
dealing with both renormalisation group running effects
and canonical normalisation corrections.
Using this formalism we have investigated the
third family wave-function corrections to the theoretical
predictions of tri-bimaximal neutrino mixing.  
We found that at leading
order both effects can be subsumed into a single universal parameter
$\eta$.  Including also the leading order Cabibbo-like charged
lepton mixing corrections, which typically arise in unified flavour
models, we have derived the theoretically stable sum rule $s=r\cos \delta
+\frac{2}{3}a$ where $s$, $r$ and $a$ parametrize
the deviations of the solar, reactor and atmospheric mixing angles
from their tri-bimaximal values 
and $\delta$ is the leptonic Dirac CP phase. 
Such a sum rule is testable in future high precision neutrino experiments
\cite{Group:2007kx}.

\section*{Acknowledgements}
We acknowledge partial support from the following grants:
PPARC Rolling Grant PPA/G/S/ 2003/00096;
EU Network MRTN-CT-2004-503369;
EU ILIAS RII3-CT-2004-506222;
NATO grant PST.CLG.980066.


\begin{thebibliography}{10}

\bibitem{HPS}
P.~F.~Harrison, D.~H.~Perkins and W.~G.~Scott,
Phys.\ Lett.\ B {\bf 530} (2002) 167;
P.~F.~Harrison and W.~G.~Scott,
Phys.\ Lett.\ B {\bf 535} (2002) 163;
P.~F.~Harrison and W.~G.~Scott,
Phys.\ Lett.\ B {\bf 557} (2003) 76.




\bibitem{sumrule}
S.~F.~King,
JHEP {\bf 0508} (2005) 105;
I.~Masina,
  Phys.\ Lett.\  B {\bf 633} (2006) 134;
S.~Antusch and S.~F.~King,
  Phys.\ Lett.\ B {\bf 631} (2005) 42;
S.~Antusch, P.~Huber, S.~F.~King and T.~Schwetz,
  JHEP {\bf 0704} (2007) 060.


\bibitem{Frampton:2004ud}
  P.~H.~Frampton, S.~T.~Petcov and W.~Rodejohann,
  Nucl.\ Phys.\  B {\bf 687} (2004) 31;
  A.~Dighe, S.~Goswami and W.~Rodejohann,
  Phys.\ Rev.\  D {\bf 75} (2007) 073023;
F.~Plentinger and W.~Rodejohann,
  Phys.\ Lett.\ B {\bf 625} (2005) 264;
R.~N.~Mohapatra and W.~Rodejohann,
  Phys.\ Rev.\ D {\bf 72} (2005) 053001;
  K.~A.~Hochmuth, S.~T.~Petcov and W.~Rodejohann,
  arXiv:0706.2975 [hep-ph].



\bibitem{Altarelli:2006kg}
  G.~Altarelli, F.~Feruglio and Y.~Lin,
  Nucl.\ Phys.\  B {\bf 775} (2007) 31;
  G.~Altarelli and F.~Feruglio,
  Nucl.\ Phys.\  B {\bf 741} (2006) 215;
  G.~Altarelli and F.~Feruglio,
  Nucl.\ Phys.\  B {\bf 720} (2005) 64.


\bibitem{Ma:2007wu}
  E.~Ma,
  arXiv:0709.0507 [hep-ph];
  E.~Ma,
  Mod.\ Phys.\ Lett.\  A {\bf 22} (2007) 101;
  E.~Ma,
  Mod.\ Phys.\ Lett.\  A {\bf 21} (2006) 2931;
   E.~Ma, H.~Sawanaka and M.~Tanimoto,
  Phys.\ Lett.\  B {\bf 641} (2006) 301;
  B.~Adhikary et al., 
  Phys.\ Lett.\  B {\bf 638} (2006) 345;
  S.~L.~Chen, M.~Frigerio and E.~Ma,
  Nucl.\ Phys.\  B {\bf 724} (2005) 423.




\bibitem{deMedeirosVarzielas:2005ax}
  I.~de Medeiros Varzielas and G.~G.~Ross,
  Nucl.\ Phys.\  B {\bf 733} (2006) 31;
  I.~de Medeiros Varzielas, S.~F.~King and G.~G.~Ross,
  Phys.\ Lett.\  B {\bf 644} (2007) 153;
  I.~de Medeiros Varzielas, S.~F.~King and G.~G.~Ross,
 Phys.\ Lett.\  B {\bf 648} (2007) 201;
  S.~F.~King and M.~Malinsky,
  Phys.\ Lett.\  B {\bf 645} (2007) 351;
  S.~F.~King and M.~Malinsky,
  JHEP {\bf 0611} (2006) 071;
  C.~Luhn, S.~Nasri and P.~Ramond,
  Phys.\ Lett.\  B {\bf 652} (2007) 27.


\bibitem{Harrison:2003aw}
  R.~N.~Mohapatra, S.~Nasri and H.~B.~Yu,
  Phys.\ Lett.\  B {\bf 639} (2006) 318;
  R.~N.~Mohapatra and H.~B.~Yu,
  Phys.\ Lett.\  B {\bf 644} (2007) 346;
  M.~C.~Chen and K.~T.~Mahanthappa,
  Phys.\ Lett.\  B {\bf 652} (2007) 34.

\bibitem{Chan:2007ng}
  A.~H.~Chan, H.~Fritzsch and Z.~z.~Xing,
  arXiv:0704.3153 [hep-ph];
 Z.~z.~Xing,
  Phys.\ Lett.\ B {\bf 618} (2005) 141;
  Shu Luo, Z.~z.~Xing, 
Phys.\ Lett.\ B {\bf 632}(2006) 341; 
  Z.~z.~Xing, Phys. Lett. B {\bf 533} (2002) 85; 
  S.~K.~Kang, Z.~z.~Xing and S.~Zhou,
  Phys.\ Rev.\  D {\bf 73} (2006) 013001;
  M.~Hirsch et al,
  Phys.\ Rev.\  D {\bf 75} (2007) 053006;
  X.~G.~He and A.~Zee,
  Phys.\ Lett.\  B {\bf 645} (2007) 427.

\bibitem{Antusch:2007jd}
  S.~Antusch, L.~E.~Ibanez and T.~Macri,
  JHEP {\bf 0709} (2007) 087.

\bibitem{King:2006hn}
As stated in the Introduction, this situation is realised in many
classes of TB mixing models.  
The precise meaning of the statement that the TB mixing comes entirely
from the neutrino sector can also be expressed in 
a basis invariant way as discussed for example in:
  S.~F.~King,
  Nucl.\ Phys.\  B {\bf 786} (2007) 52.



\bibitem{King:2007pr}
  S.~F.~King,
  arXiv:0710.0530 [hep-ph].

\bibitem{Yao:2006px}
{\bf Particle Data Group} Collaboration, W.~M. Yao {\em et~al.}, {\it Review of
  particle physics},  {\em J. Phys.} {\bf G33} (2006) 1--1232.

\bibitem{RGE}
P.~H. Chankowski and Z.~Pluciennik, 
Phys. Lett. \textbf{B316} (1993), 312--317;
K.~S. Babu, C.~N. Leung, and J.~Pantaleone, 
Phys. Lett. \textbf{B319} (1993), 191--198;
S.~F.~King and N.~N.~Singh,
  Nucl.\ Phys.\  B {\bf 591} (2000) 3; 
S.~Antusch, M.~Drees, J.~Kersten, M.~Lindner, and M.~Ratz, 
Phys. Lett. \textbf{B519} (2001)
  238--242;
%
S.~Antusch, M.~Drees, J.~Kersten, M.~Lindner, and M.~Ratz, 
Phys.\ Lett.\  B {\bf 525} (2002) 130;
%
S.~Antusch, J.~Kersten, M.~Lindner, and M.~Ratz, 
Phys. Lett. \textbf{B538} (2002),  87--95;
%
S.~Antusch and M.~Ratz, 
JHEP \textbf{07}  (2002), 059.

\bibitem{RGEanalytical}
  S.~Antusch, J.~Kersten, M.~Lindner, M.~Ratz and M.~A.~Schmidt,
  JHEP {\bf 0503} (2005) 024;
  S.~Antusch, J.~Kersten, M.~Lindner and M.~Ratz,
  Nucl.\ Phys.\  B {\bf 674} (2003) 401.
   
\bibitem{Dighe:2006sr}
  A.~Dighe, S.~Goswami and W.~Rodejohann,
  Phys.\ Rev.\  D {\bf 75} (2007) 073023.

\bibitem{King:2003xq}
S.~F. King and I.~N.~R. Peddie, 
{\em Phys. Lett.} {\bf B586} (2004) 83--94.

\bibitem{King:2004tx}
S.~F. King, I.~N.~R. Peddie, G.~G. Ross, L.~Velasco-Sevilla, and O.~Vives, 
{\em JHEP} {\bf 07} (2005) 049.

\bibitem{Antusch:2007re}
S.~Antusch, S.~F. King, and M.~Malinsky, 
 arXiv:0708.1282 [hep-ph].

\bibitem{Antusch:2004yx}
  S.~Antusch, P.~Huber, J.~Kersten, T.~Schwetz and W.~Winter,
  Phys.\ Rev.\  D {\bf 70} (2004) 097302.

\bibitem{inpreparation}
S.~Antusch, S.~F. King, and M.~Malinsky, in preparation.

\bibitem{Group:2007kx}
 The ISS Physics Working Group,
  ``Physics at a future Neutrino Factory and super-beam facility,''
  arXiv:0710.4947 [hep-ph].

 
\end{thebibliography}
\end{document}